\documentclass[twocolumn]{aastex701}

\usepackage{lipsum}
\usepackage{amsmath}
\usepackage{multirow}
\usepackage{tikz}
\usepackage{fvextra}

\begin{document}

\title{The Missing Black Hole in the Large Magellanic Cloud: A Dynamical Prediction for Its Present-Day Location}

\author[orcid=0009-0002-9263-5574,sname='Chisholm']{Robin Chisholm}
\affiliation{University of Wisconsin, Madison, Department of Physics, 1150 University Avenue, Madison, WI 53706, USA}
\email[show]{rschisholm@wisc.edu}  

\author[orcid=0000-0003-2676-8344,sname='D'Onghia]{Elena D'Onghia} 
\affiliation{University of Wisconsin, Madison, Department of Physics, 1150 University Avenue, Madison, WI 53706, USA}
\affiliation{University of Wisconsin, Madison, Department of Astronomy, 475 N. Charter Street, Madison, WI 53706, USA}
\email{edonghia@astro.wisc.edu}

\author[0000-0002-7339-3170]{Niv Drory}
\affiliation{McDonald Observatory, The University of Texas at Austin, 1 University Station, Austin, TX 78712, USA}
\email{drory@astro.as.utexas.edu}

\author[0000-0003-0724-4115]{Andrew J. Fox}
\affiliation{AURA for ESA, Space Telescope Science Institute, 3700 San Martin Drive, Baltimore, MD 21218, USA}
\email{afox@stsci.edu}

\author[0000-0002-6406-0016]{Noam Libeskind}
\affiliation{Leibniz-Institut f\"ur Astrophysik Potsdam (AIP), An der Sternwarte 16, D-14482 Potsdam, Germany}
\email{nlibeskind@aip.de}

\begin{abstract}

There are indications that the Large Magellanic Cloud (LMC) hosts a central supermassive black hole (SMBH), yet no direct detection exists: one reason is that its present-day location need not necessarily coincide with the photometric or kinematic centers. 
Here we frame a dynamical targeting problem and predict the present-day sky location of the LMC SMBH. We model the recent LMC–Small Magellanic Cloud (SMC) interaction, which induces a bar–disk offset, and integrate SMBH orbits within a realistic, time-dependent LMC–SMC–Milky Way (MW) potential. 
We show that the MW tidal field at present epoch exceeds that of the SMC by at least an order of magnitude, setting a preferred direction for coherent, vertical low-amplitude displacements, while the disk asymmetry drives in-plane offsets. 
Projecting onto the observed LMC frame, the mean of our predicted SMBH probability distribution is located at $(\alpha,\delta)=(80.23^{\circ},-69.55^{\circ})$,
with a $1\sigma$ confidence ellipse having semi-major and semi-minor axes of $(a,b)=(1.34^{\circ},0.56^{\circ})$, respectively.
This mean is $\sim6$ arcminutes north of the adopted dynamical center,
while the mode coincides with the dynamical center and lies well within the $1\sigma$ confidence region. Furthermore, all literature estimates of the LMC center considered here fall within the $2\sigma$ confidence region. 
This provides a concrete, testable target for ongoing and upcoming spectroscopic surveys, transforming the search for the LMC SMBH from an unconstrained problem into a focused observational strategy.


\end{abstract}


\section{Introduction}\label{sec:introduction} 

In the current paradigm of galaxy formation, nearly all massive galaxies are expected to host a central supermassive black hole (SMBH) \citep{lynden-bell+1969, richstone+1998}. 
Despite growing evidence for SMBHs in massive dwarf galaxies ($M_\ast \sim 10^9\,M_\odot$), and indirect indications for one in the Large Magellanic Cloud (LMC), the nearest massive dwarf galaxy, from hypervelocity stars \citep[][which constrains the mass to be $M_{\bullet}\sim6\times10^5$ M$_{\odot}$]{han+2025, lucchini+2025}, no confirmed detection exists. 
Given its proximity and mass, the LMC provides a unique opportunity to probe SMBH formation and dynamics at the low-mass end, were its putative SMBH detected.

A key difficulty is that, in a strongly perturbed system such as the LMC, the SMBH is not expected to coincide with the photometric or kinematic centers of the galaxy. 
The LMC has experienced a complex recent interaction history with the Small Magellanic Cloud (SMC) \citep{besla+2012, donghia+2016, lucchini+2021}, and is subject to a strong tidal field from the Milky Way (MW). 
These perturbations are expected to induce a displacement, or ``sloshing,'' of the SMBH within the host potential \citep{boldrini+2020}, such that the present-day location of the SMBH may be offset from any commonly adopted dynamical center. Compounding this, different tracers yield discrepant determinations of the LMC center \citep{kim+1998, vanDerMarel+2001, choi+2022, kacharov+2024, rathore+2025}, making targeted observational searches inherently uncertain.

These considerations suggest that the non-detection of a SMBH in the LMC is not solely a sensitivity issue, but fundamentally a problem of localization: where should we look? 
Previous attempts to constrain the SMBH position using hypervelocity stars trace the ejection site to $\sim 2^\circ$ north of the canonical centers \citep{lucchini+2025}, but do not uniquely determine its present-day location, particularly given the hundreds of Myrs elapsed since ejection and the subsequent dynamical evolution of the system.

Moreover, a direct prediction of the present-day location of the LMC SMBH from cosmological simulations is generally limited. 
In most large-scale simulations, SMBHs are artificially repositioned to the local potential minimum at each timestep to avoid numerical wandering \citep{springel+2005, tremmel+2015, weinberger+2020}, effectively suppressing any physically meaningful displacement relative to the host galaxy.
While necessary for numerical stability, this procedure erases precisely the dynamical information required to track off-centered SMBHs in perturbed, low-mass systems such as the LMC\footnote{Some large-scale simulations, however, have properly modeled SMBH motion via subgrid dynamical friction prescriptions \citep{sharma+2020, bird+2022}. However, even in these simulations where repositioning is relaxed, resolution limits generally prevent robust predictions of SMBH trajectories on the $\sim 10$--$100$ pc scales relevant here.}. 
As a result, the present-day phase-space position of a putative LMC SMBH cannot be uniquely determined from first principles, but must instead be treated as a probabilistic outcome shaped by the galaxy’s recent interaction history and tidal environment. 
In this sense, the problem is not to identify a single deterministic location, but to predict a probability distribution for where the SMBH is most likely to reside.

In this paper, we address this problem by developing a dynamical semi-analytic framework to predict the present-day location of a putative SMBH in the LMC. In \S\ref{sec:methodology}, we describe our methodology to construct a realistic, time-dependent model of the LMC--SMC--MW system, incorporating the recent LMC--SMC interaction that induces the observed bar--disk offset \citep{yozin+2014, pardy+2016}, and integrate the orbits of SMBH test particles within this evolving potential. 
In \S\ref{sec:analytics}, we provide complementary analytic arguments that show the present epoch tidal field induced by the MW exceeds that of the SMC by approximately one order of magnitude, setting a preferred direction for any displacement perpendicular to the disk. 
In \S\ref{sec:results}, by projecting the resulting phase-space distribution into the observational frame, we derive a probability map for the SMBH’s present-day sky location. In \S\ref{sec:discussion}, we discuss the implications of our findings. And in \S\ref{sec:conclusion}, we provide concluding remarks. This approach transforms the search for the LMC SMBH from an unconstrained problem into a targeted observational test, providing concrete guidance for ongoing and upcoming surveys.

\section{Methodology}\label{sec:methodology}

We develop a semi-analytic framework to model the dynamical evolution of a putative central SMBH in the LMC under the combined effects of internal restoring forces and external tidal perturbations, to determine its present-day spatial probability distribution.

\subsection{Equations of Motion and Orbit Integration}

The SMBH is modeled as an ensemble of test particles evolving in the combined, time-dependent LMC-SMC-MW gravitational potential.
Its motion is governed by
\begin{equation}
\ddot{\mathbf{x}} = -\nabla \Phi_{\mathrm{tot}}(\mathbf{x}, t),
\end{equation}
where
\begin{equation}
\Phi_{\mathrm{tot}} = \Phi_{\mathrm{LMC}} + \Phi_{\mathrm{MW}} + \Phi_{\mathrm{SMC}},
\end{equation}
and includes both external tidal contributions (\S\ref{sec:tidal_field}) and the internal, evolving LMC potential (\S\ref{sec:lmc_model}).
This formulation isolates the essential physics of the problem: the SMBH responds to a time-dependent, non-axisymmetric potential in which the minimum itself evolves in time. As a result, even in the absence of impulsive kicks, the SMBH can develop coherent displacements through the combined action of internal asymmetries and external tidal forcing.

We integrate the equations of motion numerically using the \texttt{galpy} framework \citep{Bovy2015}, which allows for flexible implementation of time-dependent and composite potentials. The integrations are performed at high temporal resolution ($\sim1~\rm Myr$) to accurately capture both the short-timescale response to disk asymmetries and the longer-timescale modulation induced by the external tidal field.
This framework is designed to efficiently explore the phase-space response of the SMBH across a wide range of initial conditions, enabling a robust probabilistic prediction of its present-day location.

\subsection{External Tidal Field}\label{sec:tidal_field}

The LMC evolves within the time-dependent gravitational field of the MW and SMC; these external perturbers generate differential forces across the LMC, displacing the SMBH from the instantaneous minimum of the host potential.
We model both the MW and SMC as analytic potentials and compute their contributions along the orbit of the LMC, using observationally constrained present-day phase-space coordinates \citep{zivick+2018, graczyk+2020}. 

The MW is placed at a static offset with respect to the LMC disk.
To accurately reconstruct and model the recent interaction history of the SMC relative to the LMC, we integrate backwards the orbit of the SMC, identifying a most recent pericentric passage $t_{\rm LMC-SMC}\sim 150$ Myr ago with an impact parameter of $\varrho\sim4$ kpc and a retrograde collision, consistent with previous studies \citep{besla+2012, zivick+2018}. 
The SMC then dynamically follows its orbit over the course of the integration time.

The resulting tidal field is anisotropic and time-dependent, reflecting the geometry and history of the interaction.
Crucially, it is the tidal (differential) acceleration—not the absolute gravitational force—that governs the SMBH displacement. For a SMBH near the center of the LMC, the external field introduces a differential acceleration across the LMC proportional to the gradient of the force, which competes with the internal restoring force.
This interplay produces a forced, oscillatory response: the internal potential restores the SMBH toward the center, while the external tidal field coherently alters an initially displaced orbit along preferred directions set by the large-scale environment. As shown in \S\ref{sec:analytics}, the MW tidal field at present epoch exceeds that of the SMC, defining the primary axis of displacement, although the internal disk asymmetry dominates in-plane motion.

\subsection{Gravitational Potential of the LMC}\label{sec:lmc_model}

We model the LMC as a composite, time-dependent gravitational potential consisting of three components: a dark matter halo, a stellar disk, and a central bar. The dark matter halo is represented by a spherical Hernquist profile \citep{hernquist1990}, and the stellar disk is modeled as an axisymmetric Miyamoto–Nagai component \citep{miyamoto+1975}, with parameters chosen to reproduce the observed mass and scale length of the LMC \citep{kacharov+2024}. Furthermore, we include a rotating bar modeled as a triaxial ellipsoid tuned to observations \citep{rathore+2025}, with a fixed pattern speed \citep{jimenez-arranz+2024}. 

A key ingredient of our model is the time-dependent displacement between the disk and the halo, which mimics the large-scale response of the LMC to its recent interaction with the SMC. 
This interaction induces a strong non-axisymmetric response in the LMC disk, an  excitation of an $m=1$ mode, corresponding to a coherent offset between the stellar disk and the central potential. In this picture, the bar remains anchored near the minimum of the potential (which corresponds to the center of the halo), while the disk undergoes large-scale coherent oscillations about it with amplitudes of order \noindent $\mathcal{O}(1~\rm kpc)$ \citep{yozin+2014, pardy+2016}. Naturally, we adopt the LMC bar center \citep{rathore+2025} as the dynamical center for our model.

\begin{figure*}[ht]
\epsscale{1.1}
\plotone{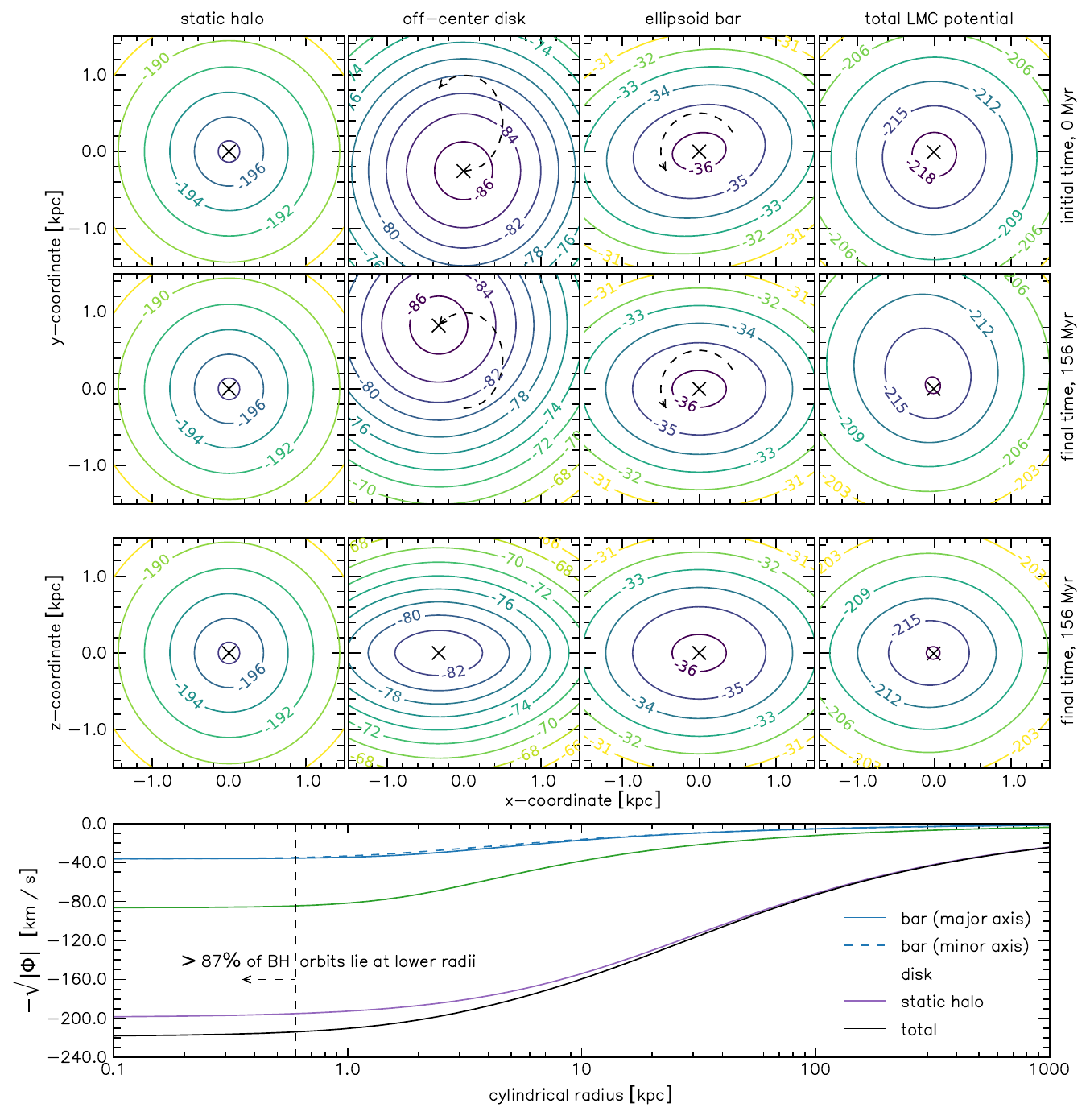}
\caption{
[\textit{top three rows}] Equipotential contour plots of $-\sqrt{|\Phi_{\rm LMC}|}$ in units of $\text{km}~\text{s}^{-1}$ of the various components of our LMC model: [\textit{far left column}] static spherical dark matter halo, modeled with a Hernquist profile, [\textit{center left column}] off-center ``wobbling" time-dependent stellar disk, modeled with a Miyamoto-Nagai disk profile, [\textit{center right column}] rotating perfect ellipsoid bar, and [\textit{far right column}] total potential model used for the LMC 
(Note: these plots do not include our MW- and SMC-models, which are incorporated for orbit integration of the finalized probability distribution). 
The first and second row show the components ``face-on", at the start of orbit integration ($t_{i}=0$ Myr) and the mean final integration time ($t_{f}=156$ Myr), respectively. The third row shows each component ``edge-on", also at the mean final integration time.
Dashed arrows in the two center columns indicate the general trajectory and direction of rotation for the off-center disk and bar. The black ``$\times$"s show the instantaneous position of the minimum potential for each component (which for the total potential, in the rightmost column, corresponds to the dynamical center).
[\textit{bottom row}] Gravitational potential profiles $-\sqrt{|\Phi_{\rm LMC}|}$ in units of $\text{km}~\text{s}^{-1}$ of the various components (including along the major and minor axes of the bar) in the equatorial plane at initial time ($t_i=0~\rm Myr$). $>87\%$ of SMBH ensemble particles ($2\sigma$ confidence ellipse extent) remain below the vertical dashed line, indicating the relevant scales.
}
\label{fig:potential}
\end{figure*}

Rather than performing a full N-body and SPH simulation, we implement this effect through a prescribed oscillatory motion of the disk potential relative to the halo center.
This “wobbling” disk potential introduces a time-dependent asymmetry that drives systematic, in-plane perturbations to SMBH orbits. Importantly, this approach isolates the dominant dynamical mechanism—an evolving, non-axisymmetric potential—while maintaining full control over the underlying physics. 
Equipotential contour plots of the various components of our LMC model, as well as their gravitational potential profiles, 
can be found in Figure \ref{fig:potential}.
All adopted model parameters for the LMC-SMC-MW system can be found in Table \ref{tab:parameters}.




\subsection{Initial SMBH Phase-Space Distribution}

Given the time-dependent potential described above, the present-day location of the SMBH depends on its phase-space configuration prior to the most recent LMC--SMC interaction. As these initial conditions are not directly constrained, we adopt a probabilistic approach, by sampling the SMBH phase space within the inner regions of the LMC. 
In low-mass systems such as the LMC, the shallow central potential and reduced efficiency of dynamical friction can allow SMBHs to remain offset from the potential minimum. 

Motivated by both observations and simulations of dwarf galaxies \citep{bellovary+2021, pfister+2019}, we therefore allow for modest initial spatial and kinematic offsets; cylindrical radii $\varrho$ are drawn from a log-normal distribution centered on $0.5~\rm kpc$, which approaches zero as $\varrho\rightarrow0$, and possesses an asymptotic tail at larger radii.
The azimuthal angle $\varphi$ is sampled uniformly, and the vertical coordinate $z$ follows a Gaussian distribution about the disk plane $(\sigma_z=100~\rm pc)$. 
Velocities are assigned assuming an isotropic dispersion consistent with measurements of the central LMC \citep[$\sigma_v\sim35~\rm km/s$, ][]{alves+2000, vanDerMarel+2001}, with an added rotational component aligned with the disk.

This ensemble spans a physically motivated range of initial configurations, allowing us to propagate uncertainty in the SMBH’s prior state into a statistical prediction for its present-day location via Monte Carlo sampling, evolved forward in time in the time-dependent potential utilizing the Chandrasekhar dynamical friction prescription \citep{chandrasekhar+1943}; we sample $10^5$ orbits.

\subsection{Construction of Probability Maps}

From the ensemble of orbit integrations, we construct a spatial probability density map for the SMBH in the reference frame of the LMC. To facilitate comparison with observations, at final integration times \footnote{
Uncertainty in the LMC--SMC collision time is incorporated by sampling final integration times from a Gaussian distribution with $\mu\pm\sigma=-156\pm15~\rm Myr$.
}, the positions of all realizations are projected into the LMC disk frame and into observable sky coordinates, transformed using a custom LMC disk coordinate frame implementation in {\tt\string astropy} \citep{astropy+2022}.

The resulting distribution is interpreted as a non-axisymmetric probability density, reflecting both the dynamical evolution of the SMBH and the uncertainty in its initial phase-space configuration. 
This procedure naturally encodes the imprint of the LMC’s interaction history, including the direction of tidal forcing and the phase of the disk oscillation.
These quantities define preferred regions for the present-day location of the SMBH, which can be directly targeted by spectroscopic and kinematic surveys.

\begin{deluxetable}{llrr}
\tabletypesize{\footnotesize}
\tablewidth{0pt} 
\tablehead{
\colhead{object} & \colhead{quantity} & \colhead{value} & \colhead{References} \\
}
\startdata 
& & & \\
 & ($\alpha,\delta$) & $(80.27,~69.65)$ deg & [1]\\
 & ($\mu_{\alpha*},\mu_{\delta}$) & $(1.88,~0.32)~\text{mas}/\text{yr}$ & \\
 \multirow{16}{4em}{LMC} & $(\iota,~\theta_{\rm nodes})$ & $(25.5,~124.0)$ deg & [2]\\
 & $v_{\rm los}$ & $264.83~\text{km}/\text{s}$ & \\
  & $d_{\text{LMC}}$ & 49.9 kpc & [3]\\
 & & & \\
 & $\Phi_{\rm halo}(r)$ & Hernquist & [4] \\
 & $M_{\rm halo}$ & $1.38\times10^{11}$ M$_{\odot}$ & [5]\\
 & $a_{\rm halo}$ & $15$ kpc & \\
 & & & \\
 & $\Phi_{\rm disk}(\rho,z)$ & Miyamoto-Nagai & [6]\\
 & $M_{\rm disk}$ & $3.5\times10^9$ M$_{\odot}$ & \\
 & $(a_{\rm disk}, b_{\rm disk})$ & $(1.5,~0.5)$ kpc & \\
 & & & \\
 & $\Phi_{\rm bar}(\rho,\phi,z)$ & Perfect Ellipsoid & \\
 & $M_{\rm bar}$ & $0.2\times M_{\rm disk}$ & \\
 & $\Omega_{\rm bar}$ & $18.5~\text{km}/\text{s}/\text{kpc}$ & [7]\\
  & & & \\
 & {$M_{\bullet}$\tablenotemark{a}} & $6\times10^5$ M$_{\odot}$ & [8]\\
 & & & \\
 & $t_{\rm LMC-SMC}$ & $-156\pm15$ Myr & \\
 & & & \\
\hline
 & & & \\
 \multirow{8}{4em}{SMC} & $(\alpha,\delta)$ & 
 $(12.54,~-73.11)$ deg & \multirow{2}{1.05em}{[9]}\\
  & $d_{\rm SMC}$ & 62.44 kpc & \\
 & ($\mu_{\alpha*},\mu_{\delta}$) & $(0.82,~-1.21)$ mas/yr & \multirow{2}{1.42em}{[10]}\\
  & $v_{\rm los}$ & 145.6 km/s & \\
 & & & \\
 & $\Phi_{\rm halo}(r)$ & Hernquist & \\
 & $M_{\rm halo}$ & $1.9\times10^{10}$ M$_{\odot}$ & \\
 & $a_{\rm halo}$ & $2.5$ kpc & \\
 & & & \\
 \hline
 & & & \\
 \multirow{6}{4em}{MW\tablenotemark{b}} & $(\alpha,\delta)$ & $(266.42,~-29.01)$ deg & [11]\\
 & $d_{\rm MW}$ & 8.178 kpc & [12] \\
 & & & \\
 & $\Phi_{\rm halo}(r)$ & NFW & [13]\\
 & $(M_{\rm 200},~c)$ & ($1.1\times10^{12}$ M$_{\odot},~10)$  & [14]\\
& & & \\
\enddata

\tablenotetext{a}{A chosen SMBH mass is necessary for dynamical friction calculations.}
\tablenotetext{b}{Sagittarius A* is adopted as the proxy for the MW center.}
\caption{Adopted model parameters; 
[1] \citet{rathore+2025}, 
[2] \citet{kacharov+2024}, 
[3] \citet{deGrijs+2014}, 
[4] \citet{hernquist1990}, 
[5] \citet{erkal+2019}, 
[6] \citet{miyamoto+1975}, 
[7] \citet{jimenez-arranz+2024}, 
[8] \citet{han+2025},
[9] \citet{graczyk+2020},
[10] \citet{zivick+2018},
[11] \citet{gordon+2023}, 
[12] \citet{grav+2019},
[13] \citet{navarro+1996},
[14] \citet{bland-hawthorn+2016}.
\label{tab:parameters}}
\end{deluxetable}

\section{Analytic Framework}\label{sec:analytics}

To interpret the numerical results and provide physical intuition, we derive an analytic estimate for the response of a low-mass SMBH to external tidal perturbations. Given the expected mass of the LMC SMBH \citep[$\lesssim 10^6\,M_\odot$,][]{boyce+2017, han+2025}, it behaves effectively as a test particle for small displacements about the center of the host potential.

We analytically approximate the central potential of the LMC as harmonic. For a mean central density $\rho_0$, the potential can be written as
\begin{equation}
\Phi(r) \simeq \frac{2\pi G \rho_0}{3} r^2,
\end{equation}
which yields the equation of motion for a displacement $x$ from the potential minimum,
\begin{equation}
\ddot{x} + \omega_0^2 x = 0,
\end{equation}
with natural oscillation frequency
\begin{equation}
\omega_0^2 = \left.\frac{\partial^2 \Phi}{\partial r^2}\right|_{r=0} = \frac{4\pi G}{3}\rho_0.
\end{equation}

\noindent This corresponds to a restoring timescale of order $\mathcal{O}(100~\rm Myr)$ for characteristic central densities of the LMC, implying that even relatively weak external perturbations can induce measurable displacements.

We now consider the effect of an external perturber of mass $m$ at distance $d$. For a SMBH displaced by $r \ll d$, the leading-order tidal acceleration is
\begin{equation}
a_{\rm tidal} \simeq 2 \frac{G m}{d^3} \, r \cos\theta,
\end{equation}
where $\theta$ is the angle between the displacement vector and the direction to the perturber. In the non-resonant, forced regime, the steady-state displacement amplitude is given by
\begin{equation}
x_{\rm max} \simeq \frac{a_{\rm tidal}}{\omega_0^2}
\simeq \frac{2 G m}{d^3 \, \omega_0^2} r
\simeq \frac{3}{2\pi \rho_0} \frac{m}{d^3} r.
\end{equation}

This scaling shows that the fractional response is set by the ratio of tidal forcing to the internal restoring density, $x/r \propto m / (\rho_0 d^3)$, and depends only weakly on the detailed internal structure of the host galaxy.

We can now compare the relative importance of the MW and the SMC as perturbers. Defining the ratio of induced displacements
\begin{equation}
R \equiv \frac{x_{\rm MW}}{x_{\rm SMC}} \sim 
\frac{M_{\rm MW} / d_{\rm MW-LMC}^3}{M_{\rm SMC} / d_{\rm SMC-LMC}^3},
\end{equation}
we find
\begin{equation*}
R \sim 10\text{--}50,
\end{equation*}
for characteristic separations $d_{\rm SMC-LMC} \sim 25~\rm kpc$ and $d_{\rm LMC-MW} \sim 50~\rm kpc$. Thus, the tidal field of the MW exceeds that of the SMC by at least an order of magnitude, setting a preferred direction for the SMBH displacement.

Using characteristic values for the LMC central density, $\rho_0 \sim 0.08\,M_\odot\,\mathrm{pc}^{-3}$, we estimate a typical displacement amplitude of
\begin{equation*}
x_{\rm max} \sim \mathcal{O}(10)\,{\rm pc},
\end{equation*}
indicating that even the dominant MW tidal field produces only modest but measurable offsets.

To provide further context regarding the tidal field acting on the putative SMBH, we compute the order of magnitude of the tidal (Jacobi) radius of the LMC within the gravitational field of the MW. 
Assuming the LMC is on an approximately circular orbit about the MW, the Jacobi radius for an orbiting body is,

\begin{equation}
r_J \simeq R_{\rm LMC} \left(\frac{M_{\rm LMC}(<r_J)}{3M_{\rm MW}(<R_{\rm LMC})}\right)^{1/3},
\end{equation}

\noindent where $M_{\rm MW}(<R_{\rm LMC})$ is the enclosed Milky Way mass at the present-day Galactocentric radius of the LMC, $R_{\rm LMC}$, which we take to be $5.4\times10^{11}~\rm M_{\odot}$ \citep{vasiliev+2019}, and $M_{\rm LMC}(<r_J)$ is the enclosed LMC mass, which for the sake of brevity, we take to be the total mass of the LMC, $1.4\times10^{11}~\rm M_{\odot}$. 
While a more detailed calculation would account for the exact orbital geometry and the enclosed LMC mass within $r_J$, these corrections modify the estimate only at the factor-of-unity level and do not affect our conclusions.
Therefore, our order-of-magnitude Jacobi radius estimate is,

\begin{equation*}
r_J \sim 20~\rm kpc.
\end{equation*}

This radius is nearly two orders of magnitude larger than the spatial scales explored by the SMBH orbits in our calculations $(\leq1~\rm kpc)$. Thus, while the Galactic tidal field can induce coherent perturbations to the SMBH motion, it is far too weak to unbind the SMBH from the LMC or substantially modify its orbit within the central regions.

This analytic framework provides a direct physical interpretation of the numerical results, linking the amplitude and direction of the SMBH displacement to the competition between internal restoring forces and external tidal perturbations.

\section{Results}\label{sec:results}
We now examine the final distribution of the SMBH ensemble, which we interpret as a probability map for the present-day location of the LMC's putative SMBH.

\subsection{Effect of ``Wobbling" Disk}

We first consider the evolution of SMBH orbits in the LMC potential alone, isolating the effect of the time-dependent disk offset induced by the LMC--SMC interaction. In this case, the dominant driver of SMBH displacement is the internal, non-axisymmetric structure of the potential.
We find that the disk ``wobbling'' produces coherent in-plane perturbations, leading to typical offsets of $\mathcal{O}(100~\rm pc)$. 
These displacements are preferentially aligned with the direction of the disk oscillation and are most pronounced for higher-energy orbits, which are more sensitive to variations in the shape of the potential well.

Despite these perturbations, the overall phase-space structure remains stable: the BH ensemble retains coherent rotation, and the velocity dispersion shows only modest evolution, with a slight reduction in the vertical component as orbits settle toward the disk plane. The offsets found in the numerical integrations ($\sim 100$ pc) relative to the analytic estimate ($\sim 10$ pc) arise from the additional contribution of the time-dependent, non-axisymmetric disk potential, which is not captured in the simple harmonic approximation.

\begin{figure*}[t]
\epsscale{1.17}
\plotone{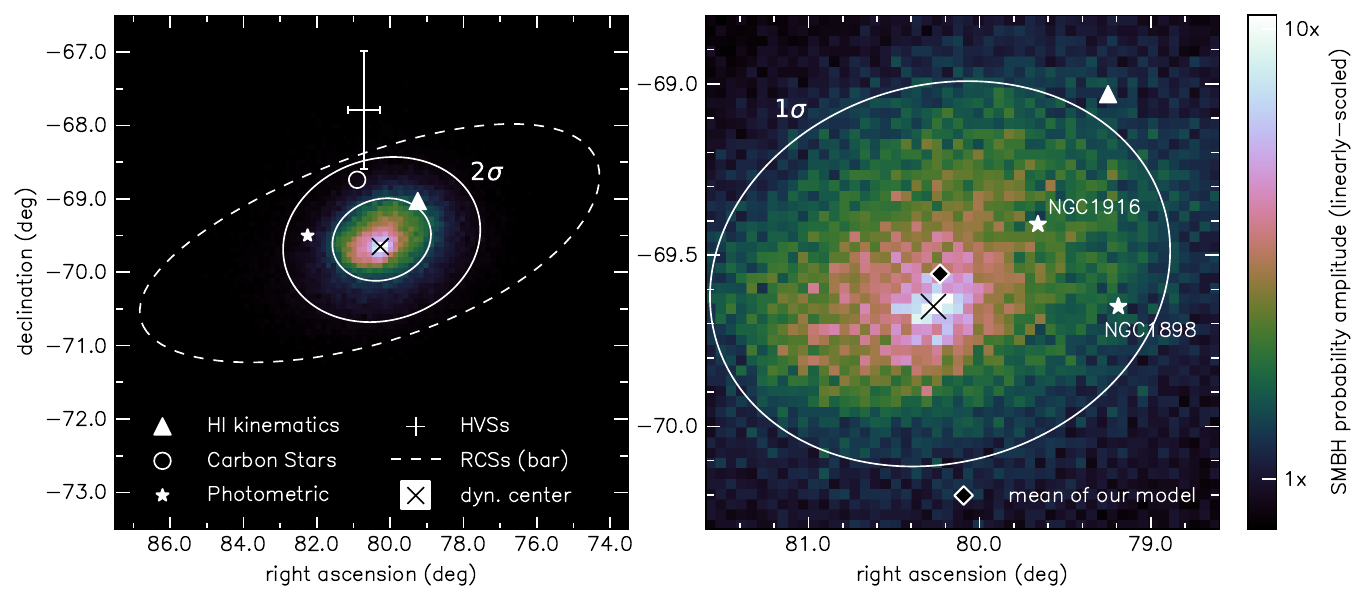}
\caption{
Predicted on-sky probability distribution for the present-day location of the putative LMC SMBH. The left panel shows the full distribution in equatorial coordinates, and the right panel \textit{zooms} into the central region. Colors indicate the relative probability density from the final SMBH ensemble. Superimposed are literature estimates of the LMC center: \ion{H}{1} \citep{kim+1998}, carbon stars \citep{wan+2020}, and photometric bar from red clump stars \citep[RCS, dashed ellipse,][]{rathore+2025} 
(this bar center is adopted as the dynamical center in our model), 
and the hypervelocity star constraints \citep{lucchini+2025}. We fit a two-dimensional Gaussian model, whose mean is displayed in the right panel with a white diamond, and $1, 2 \sigma$ confidence intervals (corresponding to $38$, $86$\%, respectively) can be seen as solid white ellipses. The right panel shows a zoomed-in version of the left, superimposed with locations of notable star clusters. The maximum of the binned map corresponds to the peak (mode) of the binned probability distribution and does not coincide exactly with the mean coordinates, instead lying at the adopted dynamical center, but still well within the $1\sigma$ contour.}

\label{fig:bhPDF}
\end{figure*}

\subsection{Effect of External Tidal Fields}

We next include the tidal influence of the MW and SMC. Consistent with the analytic expectations, the external tidal field introduces only modest additional displacements compared to the internal disk dynamics. 
The most significant effect is a systematic vertical offset of the SMBH distribution relative to the disk plane, with a mean displacement of $\sim 30$ pc. This reflects the preferred direction imposed by the MW tidal field\footnote{In the reference frame of the LMC disk, the center of MW is located approximately perpendicular below the disk.}, and introduces a coherent perturbation perpendicular to the disk. Thus, the external tidal field primarily sets the direction of the vertical response, while the disk asymmetry dominates in-plane perturbations.

The inclusion of external tides does not substantially alter the velocity dispersion, which remains at $\sigma_v \sim 40$ km s$^{-1}$. However, it does modify the bulk motion of the ensemble, introducing a slight increase in orbital eccentricity and a measurable change in the orientation of the mean velocity vector.

\subsection{Predicted On-Sky Distribution}

The final SMBH distribution is projected onto equatorial coordinates and shown in Figure \ref{fig:bhPDF}, obtaining a non-axisymmetric probability map for its present-day location. The distribution is elongated and mildly asymmetric about the adopted dynamical center,
reflecting the combined effects of internal disk asymmetries and external tidal forcing. Fitting a two-dimensional Gaussian to the projected distribution yields a mean at 

\begin{equation*}
(\alpha_0, \delta_0) = (80.23^\circ, -69.55^\circ),
\end{equation*}

and an elliptical uncertainty region ($1\sigma$ contour) with semi-major and semi-minor axes of $(a, b) = (1.34^\circ, 0.56^\circ)$. The orientation of this ellipse is aligned with the direction of the dominant tidal forcing, with a position angle of $\theta\sim 93.5
^{\circ}$ east of north.
The mean of our model lies only $\sim6~\rm arcmin$ 
from the center of the bar as found with completeness corrected red clump stars \citep{rathore+2025}, which is what we used to anchor the dynamical center of our model\footnote{
Due to the combination of the LMC's disk inclination and the tidal pull of the MW, the mean distance of the SMBH  final distribution is $\sim49.8~\rm kpc$, which is $\sim100~\rm pc$ closer to the Sun than the adopted LMC distance.
}.
Furthermore, the peak (mode) of our model lies essentially at the dynamical center, well within the $1\sigma$ confidence region. Other literature estimates of the LMC center, as shown in Figure \ref{fig:bhPDF}, are likewise within the $2\sigma$ confidence region.
Thus, our model does not provide statistically significant evidence for a present-day displacement of the SMBH from the LMC dynamical center.

\subsection{Effect of Dynamical Friction}

The mass of the SMBH enters the equations of motion only through our formulation for Chandrasekhar dynamical friction, to which the drag force is proportional to the square of the SMBH mass (aside from the Coulomb logarithm).
To quantify the impact of the dynamical friction term on our results, 
we recompute orbit integrations for SMBH ensembles with individual masses of $6\times10^{4}~\rm M_{\odot}$ and $6\times10^{6}~\rm M_{\odot}$, chosen to bracket constraints from hypervelocity stars \citep[$2.5$-$10\times10^5~\rm M_{\odot}$, ][]{han+2025}, and $6\times10^{6}~\rm M_{\odot}$ being well above the $2\sigma$ upper limit from the stellar velocity field \citep[$2.5\times10^6~\rm M_{\odot}$, ][]{boyce+2017}.

For $M_{\bullet}=6\times10^{4}~\rm M_{\odot}$, the mean of the final distribution was $(\alpha_0,\delta_0)_{6e4} = (80.22\pm1.38^{\circ}, -69.55\pm0.57^{\circ})$, $<1$ arcmin further from our adopted dynamical center than our base case, $M_{\bullet}=6\times10^5~M_{\odot}$.
For $M_{\bullet}=6\times10^{6}~\rm M_{\odot}$, the mean of the final distribution was $(\alpha_0,\delta_0)_{6e6} = (80.24\pm1.18^{\circ}, -69.57\pm0.49^{\circ})$, only $\sim1$ arcmin closer to our adopted dynamical center than the base case. Among the varied SMBH masses, the position angle of the best-fit ellipse functionally does not change.

The most significant variation in final distribution amongst the varied SMBH masses is regarding the extent of the $1\sigma$ ellipses, which notably expand with lower SMBH masses, and contract with higher SMBH masses, as expected. However, given that the mean of the final SMBH ensemble does not meaningfully vary, even with extreme variations in mass, we consider our prediction robust against adopted SMBH mass.

\section{Discussion}\label{sec:discussion}

Our predictions rely on a probabilistic treatment of the SMBH phase-space distribution prior to the most recent LMC--SMC interaction. While the adopted initial conditions are physically motivated by observations and simulations of dwarf galaxies, they remain uncertain. In low-mass systems such as the LMC, the shallow central potential and reduced efficiency of dynamical friction can allow SMBHs to remain offset from the potential minimum for extended periods. Studies suggest that such ``wandering'' SMBHs may reach offsets up to $\mathcal{O}(1~\rm kpc)$ scales \citep{bellovary+2021}, a regime only partially explored in our sampling. A broader initial spatial distribution (and increase of the initial velocity dispersion) would increase the spatial extent of the final probability map, though the preferred direction of displacement—set by the internal disk asymmetry and external tidal field—would remain unchanged. Our results should therefore be interpreted as conservative estimates  under plausible central conditions.

The mean of our predicted present-day SMBH distribution lies a few arcmins north of the LMC bar center (our adopted dynamical center), which is well within the $1\sigma$ confidence ellipse.
Furthermore, there is a $<2\sigma$ disagreement between our model's predictions and constraints derived from hypervelocity stars (HVSs), which trace an ejection site offset in a similar direction \citep{lucchini+2025}, although at a much further distance of $\sim1.9^{\circ}$.
However, the two estimates are not expected to coincide exactly. The HVS-derived position reflects the location of the SMBH at the time of ejection, several hundred Myr ago, whereas our prediction corresponds to its present-day location after subsequent dynamical evolution. The offset between the two is therefore a natural consequence of the time-dependent response of the system to internal oscillations and external tidal forcing.

If present, the LMC SMBH is likely associated with a nuclear star cluster \citep{georgiev+2016} and may exhibit observable kinematic or emission signatures. Our predicted probability map provides a concrete target region for such searches, significantly reducing the parameter space relative to previous efforts. Notably, the predicted location overlaps (i.e. with $<1\sigma$) with candidate LMC nuclear clusters (e.g., NGC~1916),
making these systems interesting targets for follow-up searches.
In addition, if the SMBH is accreting, it may be detectable through emission-line diagnostics. Ongoing surveys such as the SDSS-V Local Volume Mapper \citep{drory+2024, kollmeier+2026}, with spatial resolution of order $\mathcal{O}(10~\rm pc)$ in the central regions of the LMC, are well suited to probe this region for coronal emission lines (e.g. [\ion{Fe}{10}] $6375$\AA).

More broadly, these results demonstrate that, even in the absence of a direct detection, dynamical modeling can transform the search for SMBHs in perturbed dwarf galaxies from an unconstrained problem into a targeted observational strategy.

\section{Conclusion}\label{sec:conclusion}

We have developed a dynamical framework to predict the present-day location of a putative SMBH in the LMC. By combining a time-dependent model of the LMC--SMC--MW system with an ensemble of SMBH orbit integrations, we construct a probabilistic prediction for its spatial distribution.
A simple analytic argument shows that the response of the SMBH is governed by the competition between internal restoring forces and external tidal perturbations, with the MW dominating over the SMC 
regarding displacements perpendicular to the LMC disk. 
As a result, the SMBH is expected to undergo modest but coherent displacements, with amplitudes of order tens of parsecs, and a preferred direction set by the large-scale tidal field.
Although the external tidal field produces displacements of order tens of parsecs, while the time-dependent internal disk asymmetry can drive individual SMBH orbits to offsets of order hundreds of parsecs,
these dynamical excursions do not translate into statistically significant evidence for a present-day displacement of the overall probability distribution from the LMC center.

Our numerical integrations confirm this picture. The dominant contribution to the in-plane displacement arises from the internal, time-dependent asymmetry of the LMC disk induced by its recent interaction with the SMC, while the external tidal field introduces a systematic vertical offset and defines the overall orientation of the distribution. 
Projected onto the sky, the resulting probability map is  broad and non-axisymmetric, with its highest-probability region close to the adopted dynamical center, with no statistically significant mean offset, providing a quantitative probability region for searches.

These results transform the search for the LMC SMBH from an unconstrained problem into a predictive one. The derived probability map provides a concrete observational target for ongoing and future surveys, including integral-field spectroscopy and kinematic studies of candidate nuclear regions. More broadly, this work demonstrates that, even in dynamically complex systems where direct simulations are limited, a combined analytic and semi-analytic approach can yield robust, testable predictions for the locations of central SMBHs.

\begin{acknowledgments}
R.C. thanks the \textit{National Space Grant College and Fellowship Program} and the \textit{Wisconsin Space Grant Consortium} for their support (RFP25\_7-0). 
\end{acknowledgments}

\begin{contribution}
R.C. performed the analysis and validation and led the writing and submission of the manuscript. E.D. conceived the project and edited the manuscript. N.D. advised on observational implications. A.F. provided scientific advice and manuscript editing. N.L. provided theoretical guidance. 

\end{contribution}

\software{
{\tt\string numpy} \citep{harris+2020},
{\tt\string matplotlib} \citep{hunter+2007},
{\tt\string astropy} \citep{astropy+2022},
{\tt\string scipy} \citep{virtanen+2020}}


\bibliography{sample701}{}
\bibliographystyle{aasjournalv7}



\end{document}